%% file: EXO-11-031_temp.tex
\pdfoutput=1

\documentclass[11pt,twoside,a4paper,cmspaper,final,collab]{cms-tdr}
\def\svnVersion{175873}\def\cmsCernNo{2012-137}\def\cmsCernDate{\today}\def\cmsMessage{Submitted to Physics Letters B}

\begin{document}\cmsNoteHeader{EXO-11-031}

\hyphenation{had-ron-i-za-tion}
\hyphenation{cal-or-i-me-ter}
\hyphenation{de-vices}

\RCS$Revision: 95384 $
\RCS$HeadURL: svn+ssh://svn.cern.ch/reps/tdr2/papers/EXO-11-031/trunk/EXO-11-031.tex $
\RCS$Id: EXO-11-031.tex 95384 2012-01-16 07:51:11Z eluiggi $
\newlength\cmsFigWidth
\ifthenelse{\boolean{cms@external}}{\setlength\cmsFigWidth{\columnwidth}}{\setlength\cmsFigWidth{0.8\textwidth}}
\cmsNoteHeader{EXO-11-031} 
\title{Search for high-mass resonances decaying into $\tau$-lepton pairs in pp collisions at $\sqrt{s} = 7$\TeV}

\date{\today}

\abstract{
  A search for high-mass resonances decaying into \TT is performed using a data sample of $\textrm{pp}$ collisions
  at $\sqrt{s}=7$\TeV. The data were collected with the CMS detector at the LHC and correspond to an integrated luminosity of
  4.9\fbinv. The number of observed events is in agreement with the standard model prediction. An upper limit on the product
  of the resonance cross section and branching fraction into $\tau$-lepton pairs is calculated as a function of the resonance mass.
  Using the sequential standard model resonance $\zp_{\textrm{SSM}}$ and the superstring-inspired $E_6$ model with resonance
  $\zp_{\psi}$ as benchmarks, resonances with standard model couplings with masses below 1.4 and  1.1\TeV, respectively, are
  excluded at 95\% confidence level.
}

\hypersetup{%
  pdfauthor={CMS Collaboration},%
  pdftitle={Search for high mass resonances decaying into tau-lepton pairs in pp collisions at sqrt(s) = 7 TeV},%
  pdfsubject={CMS},%
  pdfkeywords={CMS, exotica, tau, zprime}
}

\newcommand{\emu}{\ensuremath{\tau_{\Pe}\tau_\mu}\xspace}
\newcommand{\etau}{\ensuremath{\tau_{\Pe}\tau_{\textrm{h}}}\xspace}
\newcommand{\mutau}{\ensuremath{\tau_\mu\tau_{\textrm{h}}}\xspace}
\newcommand{\ditauhad}{\ensuremath{\tau_{\textrm{h}}\tau_{\textrm{h}}}\xspace}
\newcommand{\tauhad}{\ensuremath{\tau_{\textrm{h}}}\xspace}
\newcommand{\tauell}{\ensuremath{\tau_{\ell}}\xspace}
\newcommand{\taue}{\ensuremath{\tau_{\Pe}}\xspace}
\newcommand{\taumu}{\ensuremath{\tau_\mu}\xspace}
\newcommand{\tauelltauhad}{\ensuremath{\tau_{\ell}\tau_{\textrm{h}}}\xspace}
\newcommand{\ditau}{\ensuremath{\tau^+\tau^-}\xspace}
\newcommand{\zprimeditau}{\ensuremath{\mathrm{Z}^\prime\to\tau^+\tau^-}\xspace}
\newcommand{\zditau}{\ensuremath{\mathrm{Z}\to\tau^+\tau^-}\xspace}
\newcommand{\ZTT}{\ensuremath{\mathrm{Z}\to\tau^+\tau^-}\xspace}
\newcommand{\DYTT}{\ensuremath{\mathrm{Z}\to\tau^+\tau^-}\xspace}
\newcommand{\ZEE}{\ensuremath{\mathrm{Z}\to \Pep \Pem}\xspace}
\newcommand{\ZMM}{\ensuremath{\mathrm{Z}\to \mu^+ \mu^-}\xspace}
\newcommand{\zprimeSSM}{\ensuremath{\zp_{\textrm{SSM}}}\xspace}
\newcommand{\zprimePsi}{\ensuremath{\zp_{\psi}}\xspace}

\newcommand{\ztautau}{\zditau}
\newcommand{\zee}{\ZEE}
\newcommand{\wjets}{W+jets\xspace}
\newcommand{\tauh}{\ensuremath{\tau_h}\xspace}
\newcommand{\pzeta}{\ensuremath{p_{\zeta}}\xspace}
\newcommand{\pzetavis}{\ensuremath{p_{\zeta}^{vis}}\xspace}
\newcommand{\VETslash}{\ensuremath{\overrightarrow{\ETslash}}\xspace}
\newcommand{\pzetaHat}{\ensuremath{p_{\hat{\zeta}}}\xspace}
\newcommand{\CLs}{CL$_\textrm{S}$\xspace}

\maketitle 

\section{Introduction}\label{sec:intro}
The standard model (SM) of elementary particles~\cite{glashow, weinberg, salam} provides a successful framework that describes a vast
range of particle processes involving weak, electromagetic, and strong interactions. It is widely believed, however, that the SM is
incomplete since, \eg, it fails to incorporate the gravitational force, has no dark matter candidate, and has too many parameters
whose values can not be deduced from the theory.  Many models of physics beyond the SM have been proposed to address these
problems.  A simple way of extending the SM gauge structure is to include an additional U(1) group, which requires an associated
neutral gauge boson, usually labeled as  \zp\,\cite{langacker,leike, cvetic, diener}. Although most studies have assumed
generation-independent gauge couplings for \zp, models in which the \zp couples preferentially to third generation fermions have also
been proposed\,\cite{diener,chivukula,plumacher,malkawi}. The sensitivity of the traditional searches for $Z^\prime$ production using
\EE and \MM final states may be substantially reduced in such non-universal scenarios, motivating the exploration of other allowed
final states. The sequential SM (SSM) includes a neutral gauge boson, \zprimeSSM, with the same couplings to quarks and leptons as
the SM \Z boson. Although not gauge invariant, this model has been traditionally considered by experiments studying high-mass
resonances. Other models, such as the superstring-inspired $E_{6}$ model, have more complex group  structures,  $E_{6}\to
SO(10)\times U(1)_{\psi}$, with a corresponding neutral gauge boson denoted as \zprimePsi\,\cite{hewett}.   In this Letter we present
a search for heavy resonances that decay into pairs of $\tau$ leptons using the \zprimeSSM and \zprimePsi models  as benchmarks. A
previous search reported by the CDF collaboration has ruled out a \zprimeSSM decaying into \ditau with SM couplings with mass below
399\GeV\,\cite{acosta}.

About one-third of the time $\tau$ leptons decay into lighter charged leptons plus neutrinos, whereas the other two-thirds of the
time $\tau$ leptons decay into a hadronic system with one, three, or five charged mesons, which can be accompanied by neutral pions
in addition to the $\tau$ neutrino. We refer to the leptonic and hadronic decay channels as \tauell ($\ell=e,\mu$) and \tauhad,
respectively. Four \ditau final states, \emu, \etau, \mutau, and \ditauhad, are studied using a sample of proton-proton collisions
at $\sqrt{s}=7$\TeV recorded by the Compact Muon Solenoid (CMS) detector at the Large Hadron Collider (LHC). The data sample
corresponds to an integrated luminosity of $4.94 \pm 0.11\fbinv$.

\section{The CMS Experiment}\label{sec:cms}

The central feature of the CMS apparatus is a superconducting solenoid, of 6~m internal diameter, providing a field of 3.8\,T. Within
the field volume are the silicon pixel and strip tracker, the lead-tungstate (PbWO$_4$) crystal electromagnetic calorimeter (ECAL),
and the brass/scintillator hadron calorimeter (HCAL). Muons are measured with detection planes made using three technologies: drift
tubes, cathode strip chambers, and resistive plate chambers. Extensive forward calorimetry complements the coverage provided by the
barrel and endcap detectors.

CMS uses a right-handed coordinate system, with the origin at the nominal interaction point, the $x$ axis pointing to the
center of the LHC ring, the $y$ axis pointing up (perpendicular to the plane of the LHC ring), and the $z$ axis along the
counterclockwise-beam direction. The polar angle, $\theta$, is measured from the positive $z$ axis and the azimuthal angle,
$\phi$, is measured in the $x$-$y$ plane.

The inner tracker measures charged particles within the pseudorapidity range $|\eta| < 2.5$, where
$\eta = -\ln[\tan(\theta/2)]$; muons are measured within $|\eta|< 2.4$.

A more detailed description of the CMS detector can be found elsewhere~\cite{CMS}.

\section{Lepton Reconstruction and Identification}\label{sec:leptonRecoId}

Electrons are reconstructed by combining clusters in the ECAL with tracks in the inner tracker fitted with a Gaussian sum filter
algorithm\,\cite{electronTrigger}. Electron candidates are required to have good energy-momentum  and spatial match between the
ECAL cluster and the inner track. In addition, the ratio of the energies measured in HCAL and ECAL must be consistent with an
electron signature.

Muons are reconstructed by matching hits found in the muon detectors to tracks in the inner tracker\,\cite{muonTrigger}. Quality
requirements, based on the minimum number of hits in the silicon, pixel, and muon detectors, are applied to suppress backgrounds
from punch-through and decay-in-flight pions. A requirement on the maximum transverse impact parameter with respect to the
beamspot (2\mm) largely reduces the contamination from cosmic-ray muons.

A particle-flow (PF) technique\,\cite{PFTausReco} is used for the reconstruction of hadronically decaying $\tau$ candidates. In the
PF approach, information from all subdetectors is combined to reconstruct and identify all final-state particles produced in the
collision. The particles are classified as either charged hadrons, neutral hadrons, electrons, muons, or photons. These particles
are used to reconstruct \tauhad candidates using the hadron plus strip (HPS) algorithm\,\cite{TauID}, which is designed to optimize
the performance of \tauhad identification and reconstruction by considering specific $\tau$-lepton decay modes.

\section{Signal and Background MC Samples}\label{sec:samples} Signal and background Monte Carlo (MC) samples are produced with the
\PYTHIA 6.4.22\,\cite{PYTHIA} and \MADGRAPH\,\cite{Madgraph} generators using the Z2 tune\,\cite{Z2} and the CTEQL1 parton
distribution function (PDF) set\,\cite{CTEQL1}. The \TAUOLA package\,\cite{TAUOLA} is used to decay the generated $\tau$ leptons.
All generated objects are input into a detailed {\GEANTfour}\,\cite{Geant} simulation of the CMS detector.

The \zprimeSSM and \zprimePsi signal samples are generated with seven different masses: 350, 500, 750, 1000, 1250, 1500, and
1750\GeV. Their corresponding cross sections are calculated to leading order.

The most important sources of background are the irreducible Drell--Yan process (\DYTT), production of W bosons in association with
one or more jets (\wjets), \cPqt{}\cPaqt, dibosons (WW, WZ), and QCD multijet production. Although the \ZTT background peaks around
the \Z mass, its tail extends to the region where a high-mass resonance might present itself. The \wjets events are characterized by
an isolated lepton from the decay of the W boson and an uncorrelated jet misidentified as  a light lepton or a \tauhad. Background
from \ttbar events is usually accompanied by one or two b jets, in addition to genuine, isolated leptons or \tauhad. Background from
diboson events produces both genuine, isolated leptons, when the gauge bosons decay leptonically, and a misidentified \tauhad when
they decay hadronically. Finally, QCD events are characterized by non-collimated jets with a high-multiplicity of particles, which
can be misidentified as charged light leptons and \tauhad.

\section{Event Selection}\label{sec:selections}
A new heavy neutral gauge boson decaying into $\tau$ pairs is characterized by two high-\pt, oppositely charged,
isolated, and almost back-to-back (in the transverse plane) $\tau$ candidates.

CMS uses a two-level trigger system consisting of the so called level one (\Lone) and the high level trigger (HLT). The events
selected for this analysis are required to have at least two trigger objects: an electron and a muon, a \tauhad candidate and a light
charged lepton, or two \tauhad candidates for the \emu, \tauelltauhad, and \ditauhad final states, respectively. Details of the
electron and muon triggers can be found in Refs.~\cite{electronTrigger,muonTrigger}. The \tauhad trigger algorithm requires presence
of jets reconstructed at \Lone using a 3x3 combination of calorimeter trigger  regions. Those jets are considered  as seeds for the
HLT where a simplified version of the PF algorithm is used to build the HLT \tauhad candidate. The HLT \tauhad four-momentum is
reconstructed as a sum of the four-momenta of all particles in the jet with pT above 0.5 \GeV in a cone of radius $\DR =
\sqrt{(\Delta\eta)^2 + (\Delta\phi)^2} = 0.2$ around the direction of the  leading particle.

The event selection requirements are optimized, for each individual final state, to maximize the sensitivity reach of the search.
Events are required to have at least two $\tau$ candidates satisfying the \pt, $\eta$, and isolation requirements presented in
Table~\ref{tab:leptonId}.  The isolation requirement is the strongest discriminator between genuine $\tau$ candidates and those from
misidentified QCD jets. The leptonic $\tau$ candidates (\taue, \taumu) are required to pass both track and ECAL isolation requirements.
Track isolation (TrkIso) is defined as the sum of the \pt of the tracks, as measured by the tracking system, within an isolation cone 
of radius \DR~centered around the charged light lepton track. Similarly, ECAL isolation (EcalIso) measures the amount of energy
deposited in the ECAL within the isolation cone. In both cases the contribution from the charged light lepton candidate is removed from
the sum. For \tauhad candidates, the HPS algorithm provides three isolation definitions. The ``loose" $\tau$ definition rejects a
$\tau$ candidate if one or more charged hadrons with $\pt \geq 1.0\GeV$ or one or more photons with transverse energy $\et \geq
1.5\GeV$ are found within the isolation cone. The ``medium" and ``tight" definitions require no charged hadrons or photons with \PT
greater than 0.8 and 0.5\GeV within the isolation cone, respectively. Additionally, \tauhad candidates are required to fail electron
and muon requirements. 

Pairs are formed from oppositely charged $\tau$ candidates with $\DR > 0.7$. In addition, a back-to-back requirement on the $\tau$
pairs is imposed by selecting candidates with   $\cos\Delta\phi(\tau_{1},\tau_{2}) < -0.95$,  where $\Delta\phi(\tau_{1},\tau_{2})$ is
the difference in the azimuthal angle between the $\tau$ candidates. The presence of neutrinos in the final state precludes a full
reconstruction of the mass of the \ditau system. We use the visible $\tau$-decay products and the missing transverse energy, \MET,
defined as the magnitude of the negative of the vector sum of the transverse momentum of all PF objects in the  event\,\cite{CMSMet},
to reconstruct the effective visible mass:
\ifthenelse{\boolean{cms@external}}{
\begin{multline}
   M(\tau_{1},\tau_{2},\MET) = \\
   \sqrt{(E_{\tau_{1}}+E_{\tau_{2}}+\MET)^{2}-
   (\overrightarrow{p_{\tau_{1}}}+\overrightarrow{p_{\tau_{2}}}+\overrightarrow{\MET})^{2}}.
\label{eq:ditaumass}
\end{multline}
}{
\begin{equation}
   M(\tau_{1},\tau_{2},\MET) = \sqrt{(E_{\tau_{1}}+E_{\tau_{2}}+\MET)^{2}-
   (\overrightarrow{p_{\tau_{1}}}+\overrightarrow{p_{\tau_{2}}}+\overrightarrow{\MET})^{2}}.
\label{eq:ditaumass}
\end{equation}
}
When compared with the mass obtained by using only the visible decay products of the \ditau system, the effective mass provides better
discrimination against backgrounds. The width and central value of the effective visible mass distribution, which is offset from the
true resonance mass,  depend on the true mass of the new resonance. The width-to-mass ratio of the effective mass distribution
reconstructed using Eqn.\,\ref{eq:ditaumass} varies from $\sim 25\%$ for a \zp with a generated mass of 350\GeV to $\sim 50\%$ for a
\zp with generated mass of 1750\GeV.

\begin{table*}[htbp]
  \begin{center}
    \caption{Lepton \pt, $\eta$, and isolation requirements for each \ditau final state.}
    \begin{tabular}{|l|l|c|c|c|c|}
    \hline\hline
    \multicolumn{2}{|c|}{Selection}		       	& \emu  & \etau   & \mutau & \ditauhad	\\ \hline
    \multirow{4}{*}{\taue} 	& Min \pt (\GeV) 	& 15    & 20	 & --	  & --  	  \\
    				& Max $|\eta|$ 	& 2.1   & 2.1	& --     & -- 	  \\
    				& Iso \DR 	       	& 0.3   & 0.4	 & --	  & --  	  \\
				& Max TrkIso (\GeV)	& 3.5   & 3.0	 & --	  & --  	  \\
				& Max EcalIso (\GeV) 	& 3.0   & 4.5	 & --	  & --  	  \\\hline
    \multirow{4}{*}{\taumu} 	& Min \pt (\GeV) 	& 20    & --	 & 20	  & --  	  \\
    				& Max $|\eta|$		& 2.1   & --	 & 2.1    & -- 	  \\
    				& Iso \DR 	       	& 0.5   & --	 & 0.5    & --  	  \\
				& Max TrkIso (\GeV)    	& 3.5   & --	 & 1.0    & --  	  \\
				& Max EcalIso (\GeV)  	& 3.0   & --	 & 1.0    & --  	  \\\hline
    \multirow{2}{*}{\tauhad} 	& Min \pt (\GeV)	& --    & 20	 & 20	  & 35  	  \\
    				& Max $|\eta|$		& --    & 2.1	 & 2.1    & 2.1 	  \\
    				& Iso \DR 	       	& --    & 0.5	 & 0.5    & 0.5 	  \\
    				& Iso definition 	& --    & Medium  & Medium & Loose	  \\
    \hline\hline
    \end{tabular}\label{tab:leptonId}
  \end{center}
\end{table*}

Further selection requirements are applied to suppress background contributions. To discriminate against QCD jets, \MET is
required to be greater than 20\GeV for the \emu final state and greater than 30\GeV for the \etau, \mutau, and \ditauhad  final
states. Furthermore, we consider only one-prong $\tauhad$ candidates in the \ditauhad final state.

To reduce the contamination from collisions in which W bosons are produced, events are required to  be consistent with the signature
of a particle decaying into two $\tau$ leptons. We define a unit vector ($\hat{\zeta}$) along the bisector defined by the \pt vector of
the tau candidates, and two projection variables\,\cite{CDFMSSM}:

\begin{equation}
  p_{\zeta}^{vis} = \overrightarrow{p}_{\tau_{1}}^{vis} \cdot \hat{\zeta} + \overrightarrow{p}_{\tau_{2}}^{vis} \cdot \hat{\zeta},
\label{eq:zeta}
\end{equation}

\begin{equation}
  p_{\zeta} = p_{\zeta}^{vis} + \overrightarrow{\MET} \cdot \hat{\zeta}.
\label{eq:zetavis}
\end{equation}

We require that $(1.25~\times~p_{\zeta}^{vis}) - p_{\zeta} < 10\GeV$ for the \emu final state and
$(0.875~\times~p_{\zeta}^{vis}) - p_{\zeta} < 7\GeV$, for the \mutau, \etau, and \ditauhad  final states.
Hereafter, we will refer to the inequalities based on the $p_{\zeta}$~and~$p_{\zeta}^{vis}$ variables as the \pzetaHat
requirements.

Any remaining contribution from \ttbar events is  minimized by selecting events where none of the jets has been identified as a b
jet. A jet is identified as a b jet if it has at least two tracks with impact parameter significance, defined as the ratio
between the impact parameter and its estimated uncertainty, greater than 3.3\,\cite{CMS_PAS_BTV_10-001}. In order to further reduce
the \ttbar contribution in the \emu final state, an additional requirement is imposed on the difference in the azimuthal angle between the
highest-\pt (leading) lepton ($\ell^{\textrm{lead}}$) and the \MET vector ($\cos\Delta\phi(\ell^{\textrm{lead}},\MET) < -0.6$).

Table~\ref{table:acceptance} summarizes the signal selection efficiency after all requirements have been applied for various
\zprimeSSM  masses. The uncertainties in the selection efficiencies are statistical only. The \zprimePsi model has comparable
efficiencies.

\begin{table*}[htbp]
  \caption{Signal selection efficiency for \zprimeSSM decaying into \emu, \etau, \mutau, and \ditauhad final states.}
  \centering{
  \begin{tabular}{| l | c | c | c | c |}
  \hline\hline
       \zprimeSSM Mass (\GeV)	& \emu		   	& \etau		 	& \mutau		& \ditauhad \\ [0.5ex]\hline
       350  		     	& $0.142 \pm 0.007$   	& $0.050 \pm 0.004$   	& $0.08 \pm 0.01$	& $0.014 \pm 0.001$ \\
       500  		     	& $0.199 \pm 0.008$   	& $0.064 \pm 0.005$   	& $0.13 \pm 0.01$     	& $0.022 \pm 0.001$ \\
       750  		     	& $0.257 \pm 0.008$   	& $0.083 \pm 0.007$   	& $0.15 \pm 0.01$     	& $0.030 \pm 0.001$ \\
       1000 		     	& $0.281 \pm 0.009$   	& $0.087 \pm 0.007$   	& $0.17 \pm 0.01$     	& $0.034 \pm 0.001$ \\
  \hline
  \hline
  \end{tabular}
  }
  \label{table:acceptance}
\end{table*}

\section{Background Estimation}\label{sec:extraction}
The estimation of the background contributions in the signal region is derived from data wherever possible. The general
strategy is to modify the standard selection requirements to select samples enriched with background events. These control
regions are used to measure the efficiencies for  background candidates to pass the signal selection requirements. In cases
where the above approach is not feasible, data-to-MC scale factors, defined as a ratio of efficiencies, are used to correct
the expected contributions obtained from the simulation samples.

The number of QCD events in the signal region for the \ditauhad and \tauelltauhad final states is estimated from a sample of
like-sign $\tau\tau$ candidates scaled by the opposite-sign to like-sign ratio ($R_{OS/LS}$) observed in the data. For the \ditauhad
final state, $R_{OS/LS}$ is measured for events with transverse mass  $M_{T}(\tauhad^{\textrm{lead}},\MET)$ between 15 and 90\GeV,
where $\tauhad^{\textrm{lead}}$ is the highest-\pt \tauhad candidate  and  $M_{T}(\tauhad^{\textrm{lead}},\MET) =
\sqrt{2\pt^{\tauhad^{\textrm{lead}}}\MET(1-\cos\Delta \phi)}$.  For the \tauelltauhad final states $R_{OS/LS}$ is measured from a
sample in which the muon TrkIso is between 4 and 15\GeV.  For the \emu final state, the QCD background is estimated using a sample
of non-isolated electrons and muons. The isolation efficiency, needed to
extrapolate into the signal region, is measured from a sample of like-sign candidates.

An enhanced sample of \wjets events is obtained by removing the $\Delta\phi(\tau_{1},\tau_{2})$ and \pzetaHat requirements
from the standard selections. Further enhancement of W events is obtained by requiring that the transverse mass of the lepton
and \MET system to be between 50 and 100\GeV. The number of \wjets events in the signal region is
estimated from the number of events in the \wjets enhanced sample passing the $\Delta\phi(\tau_{1},\tau_{2})$ and \pzetaHat
requirements divided by the efficiency of the transverse mass requirement measured from a sample of events passing the
complement of both the $\Delta\phi(\tau_{1},\tau_{2})$ and \pzetaHat requirements.

A high-purity sample of \ttbar events is obtained by requiring the presence of at least one reconstructed b jet with
$\pt > 20$\GeV and removing the $\Delta\phi(\tau_{1},\tau_{2})$ and \pzetaHat requirements. The \ttbar contribution in the
signal region is estimated from the number of events with one or more b jets passing the $\Delta\phi(\tau_{1},\tau_{2})$
and \pzetaHat requirements multiplied by the ratio of events passing the zero b-jet requirement and those events with one
or more b jets.  This ratio is measured from a sample with at least two additional jets in the events, which is
dominated by \ttbar events.

Control samples dominated by \ZEE and \ZMM background events are obtained by removing the \MET requirement and requiring the
$\tau$ candidates to be compatible with charged light-lepton signatures.  The number of events in each control sample is compared with
the expected contributions as determined from the simulation and are found to be in agreement. Since the \ZEE and \ZMM
backgrounds are well described by the simulation, the estimated background contribution from these sources is taken directly from
the number of simulated events passing the standard selection requirements normalized to the recorded integrated luminosity and
the next-to-next-leading order cross-section value.

The contamination from diboson production is estimated directly from the number of simulated events that pass the analysis
requirements normalized to the integrated luminosity with next-to-leading order (NLO) cross-section values.

Finally, high-purity samples of $\mathrm{Z}\to\tauelltauhad$ events in which the data-to-MC scale factor can be evaluated
are obtained by removing the \MET selection and requiring that the \tauell transverse momentum be less than 40\GeV.
Contamination from  \wjets events in these samples is reduced after requiring $M_T(\tauell,\MET) < 40\GeV$.
For the \emu final state the data-to-MC scale factor is evaluated for $M(\tau_e,\tau_\mu,\MET) < 150\GeV$.

\begin{figure*}
  \begin{center}
    \includegraphics[angle=0,width=.48\textwidth]{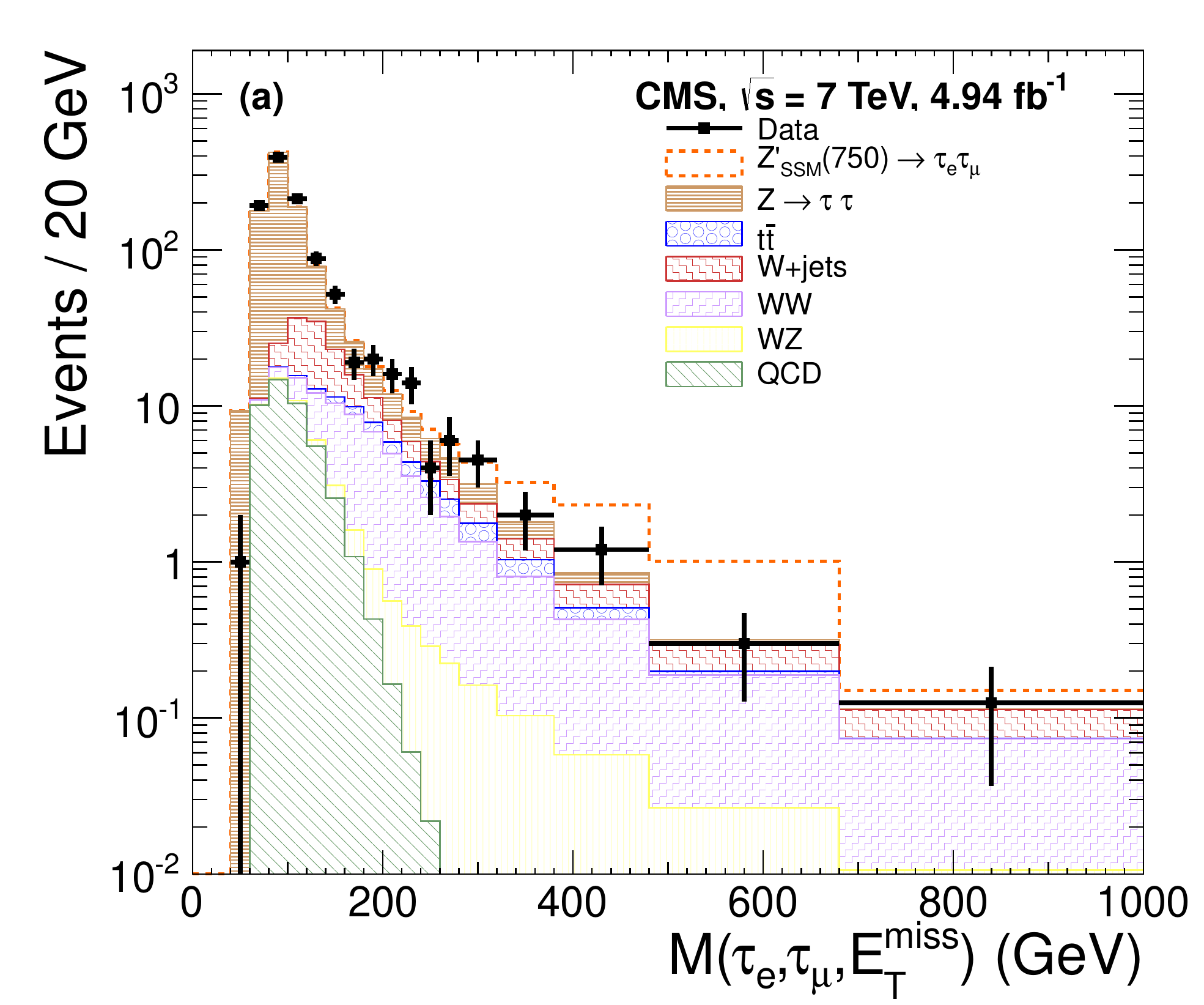}
    \includegraphics[angle=0,width=.48\textwidth]{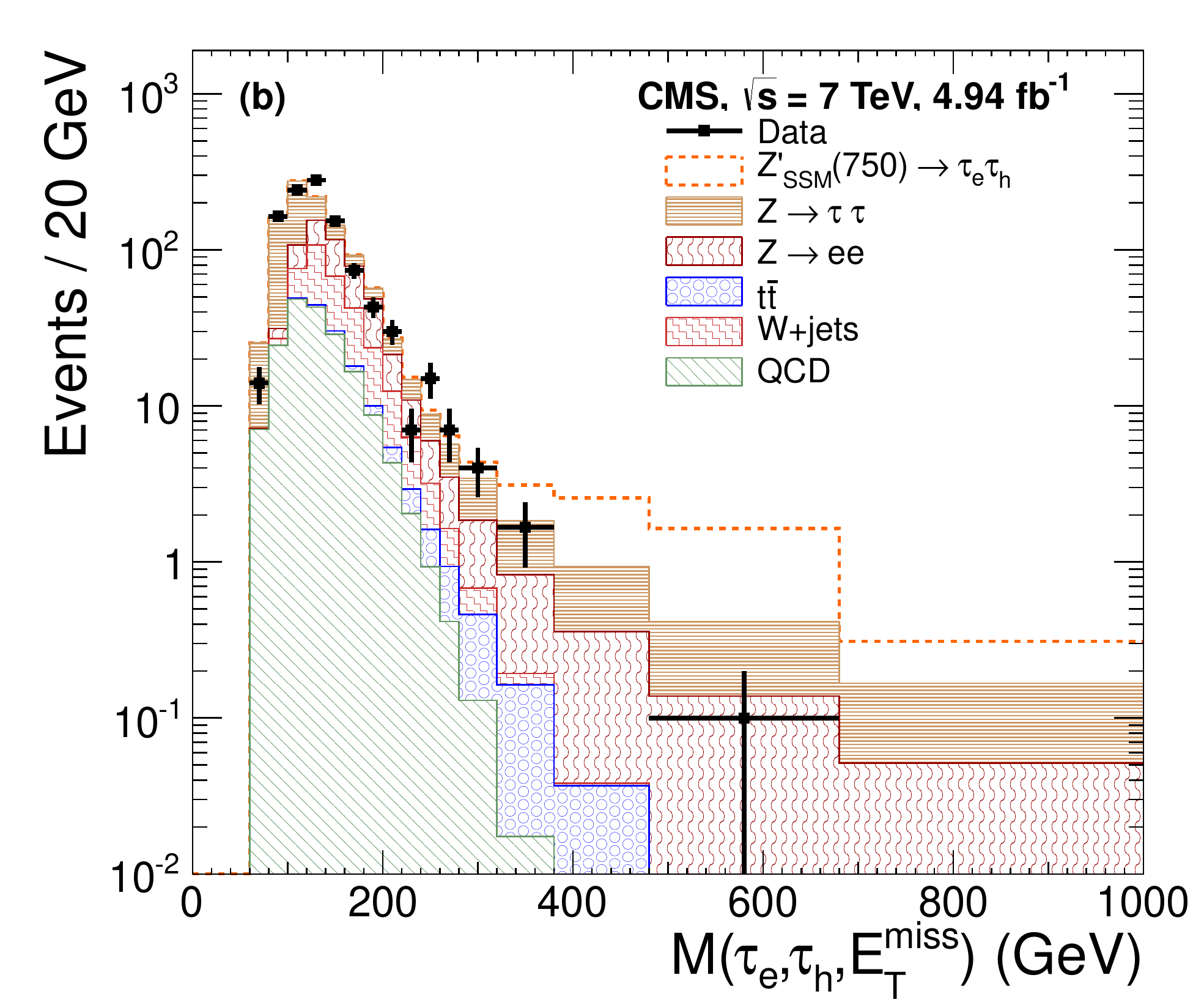}
    \includegraphics[angle=0,width=.48\textwidth]{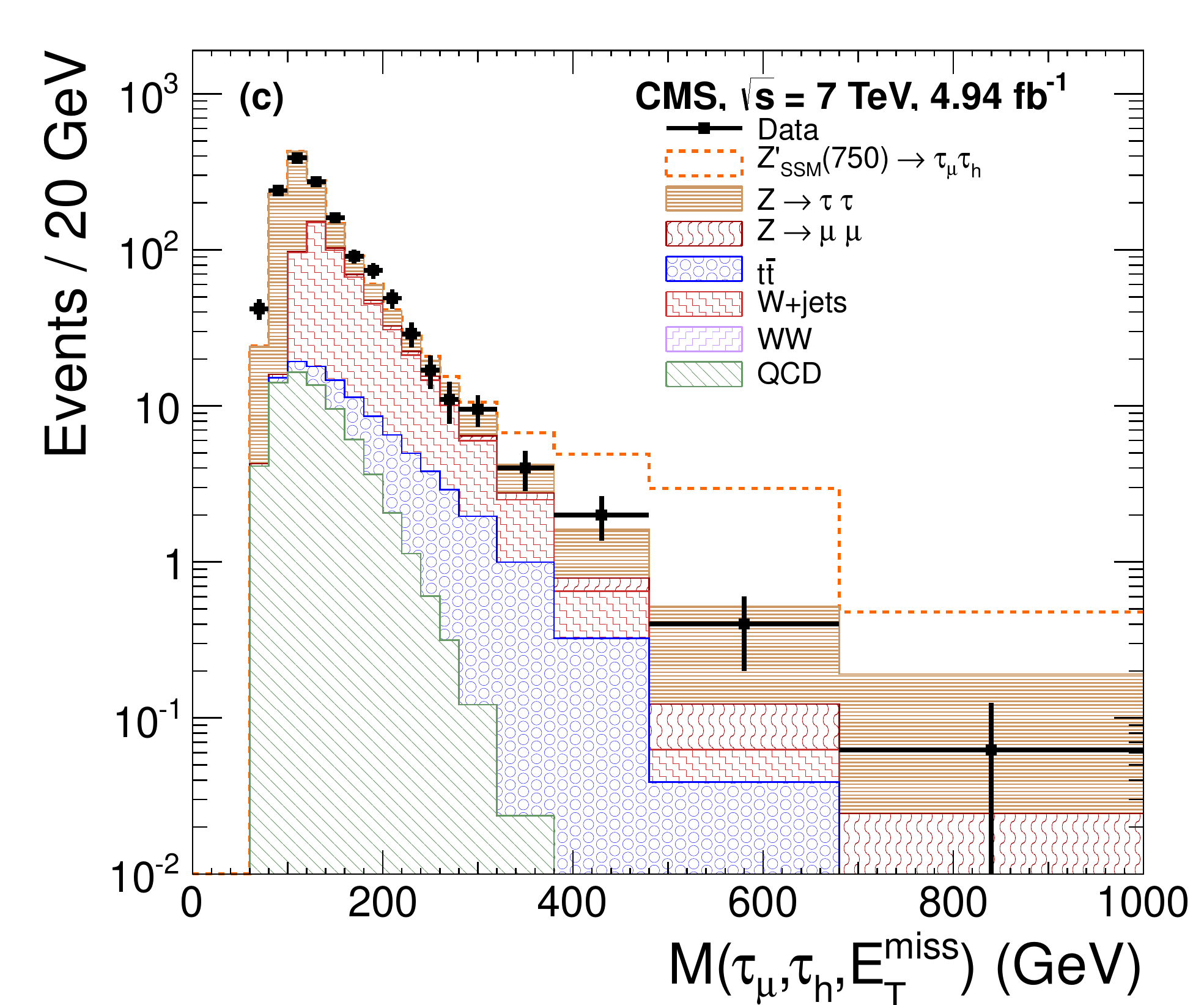}
    \includegraphics[angle=0,width=.48\textwidth]{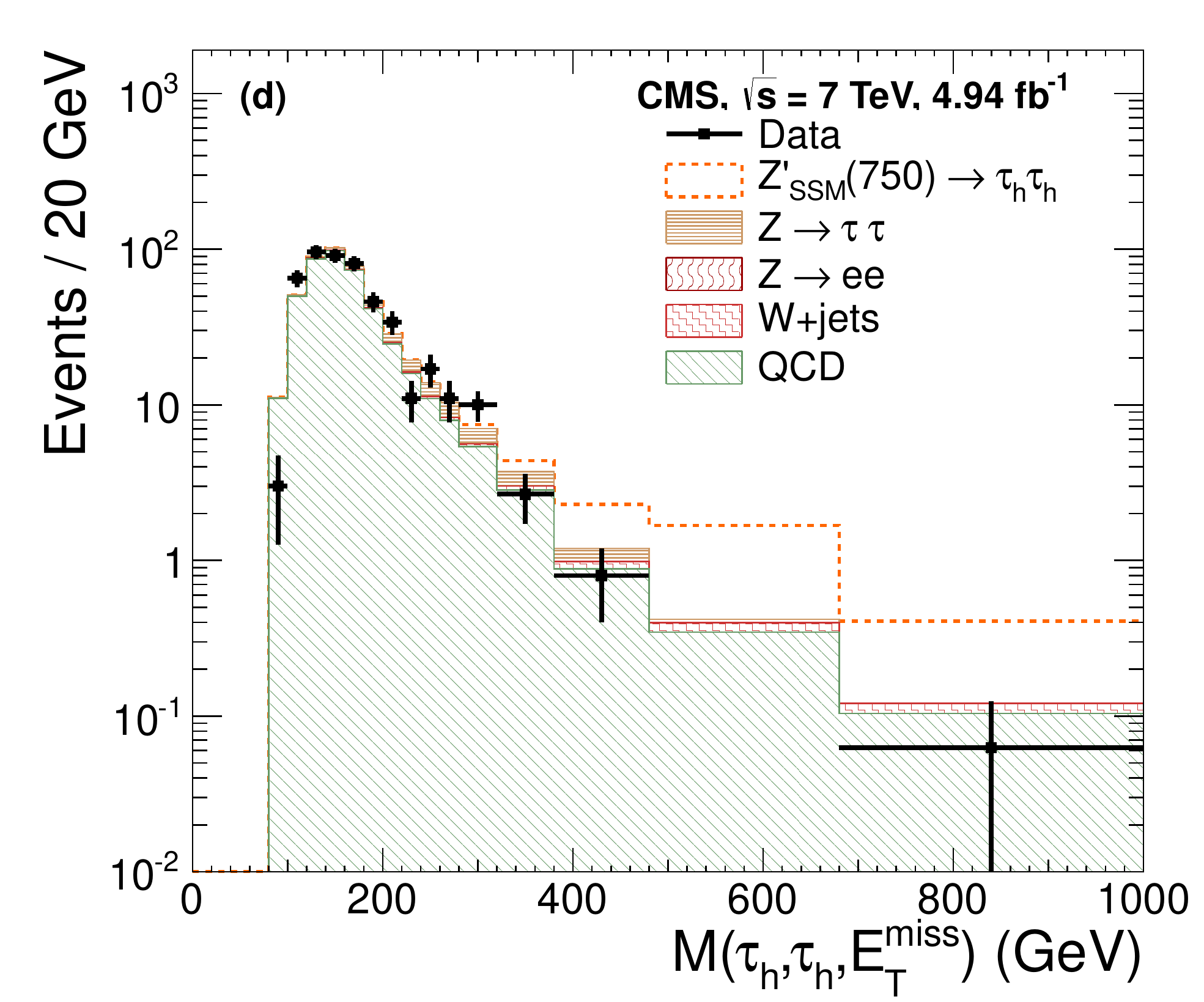}
    \caption{$M(\tau_{1},\tau_{2},\MET)$ distributions for all four final states: (a) \emu, (b) \etau, (c) \mutau, and (d) \ditauhad.
    The dashed line represents the mass distribution for a  $\zprimeSSM\to\ditau$ with a mass of 750\GeV.}
    \label{fig:masses}
  \end{center}
\end{figure*}

Table~\ref{table:expectations} lists the number of estimated background events compared with the total number of observed events in
data for each final state. The statistical and systematic uncertainties are quoted separately. Only the uncertainties on the cross
section and background estimation methods are included in the systematic uncertainties. Other effects, such as the uncertainty of
the luminosity measurement, are not included in Table~\ref{table:expectations}. The \ditau effective visible mass distributions,
$M(\tau_1,\tau_2,\MET)$, for all four final states are shown in Fig.\,\ref{fig:masses}.  The largest background sources are from
W+jet(s) and Drell--Yan production for \emu and \tauelltauhad final states, and QCD processes for \ditauhad.  We use the background
shapes normalized to the values obtained from the background estimation methods to search for a broad enhancement in the
$M(\tau_{1},\tau_{2},\MET)$ spectrum consistent with the production of a high-mass resonance state. The background shapes are taken
from the MC simulation and parametrized to extrapolate to the high-mass regions.

\begin{table*}[htbp]
  \caption{Number of observed events in data and estimated background events for the whole mass range. The first and
  second uncertainties are the statistical and systematic, respectively.}
  \centering{
  \begin{tabular}{| l | c | c | c | c |}\hline\hline
    Process  & \emu			& \etau		     	& \mutau			& \ditauhad \\ [0.5ex] \hline
    \zditau  & $816 \pm 58 \pm 44$	& $462 \pm 56 \pm 24$   & $804 \pm 53 \pm 44$   	& $30.9 \pm 3.6 \pm 4.3$	\\
    \ZMM     & --			& --		     	& $20.8 \pm 8.3 \pm 1.1$	& --				\\
    \ZEE     & --			& $220 \pm 24 \pm 11$   & --		       		& $0.66 \pm 0.33 \pm 0.22$      \\
    \wjets   & $83 \pm 15 \pm 7$	& $181 \pm 36 \pm 13$	& $459 \pm 26 \pm 29$	   	& $5.8 \pm 1.7 \pm 1.1$	   \\
    WW       & $55.6 \pm 1.4 \pm 1.9$	& --			& $24.59 \pm 0.80 \pm 0.80$   	& -- 			   \\
    WZ       & $5.60 \pm 0.35 \pm 0.22$ & --			& --			   	& -- 			   \\
    \ttbar   & $9.6 \pm 1.2 \pm 0.7$	& $10.8 \pm 2.8 \pm 0.9$& $46.2 \pm 6.9 \pm 3.7$   	& $0.00^{+0.76 + 0.15}_{-0.00}$	\\
    QCD      & $45.1 \pm 3.3 \pm 9.0$	& $185 \pm 31 \pm 19$	& $72 \pm 18 \pm 8$	   	& $467 \pm 26 \pm 67 $	   \\
    Total    & $1015 \pm 60 \pm 45$	& $1058 \pm 77 \pm 35$  & $1427 \pm 63 \pm 53$     	& $504 \pm 26 \pm 67$	   \\\hline
    Observed & $1044$			& $1043$		& $1422$		   	& $488$\\
  \hline\hline
  \end{tabular}
  }
  \label{table:expectations}
\end{table*}

\begin{table*}[htb]
  \caption{Number of observed events in data and estimated background events obtained from the integration of the
  $M(\tau_1,\tau_2,\MET)$ distribution for masses above 300\GeV. The statistical and systematic uncertainties have
  been added in quadrature.}
  \centering{
  \begin{tabular}{| l | c | c | c | c |}\hline\hline
    Process  & \emu			& \etau		     	& \mutau		& \ditauhad \\ [0.5ex] \hline
    \zditau  & $2.76 \pm 0.25$ 		& $12.5 \pm 1.7$	& $18.0 \pm 1.5$  	& $3.41 \pm 0.62$    	\\
    \ZMM     & -- 			& --			& $3.0 \pm 1.2$  	& --    		\\
    \ZEE     & -- 			& $6.94 \pm 0.83$	& --  			& --    		\\
    \wjets   & $4.86 \pm 0.97$ 		& $0.23 \pm 0.05$	& $9.58 \pm 0.81$  	& $1.83 \pm 0.69$	\\
    WW       & $8.53 \pm 0.36$ 		& --			& $4.01 \pm 0.18$ 	& --			\\
    WZ       & $1.34 \pm 0.10$ 		& --			& --  			& --			\\
    \ttbar   & $1.57 \pm 0.23$  	& $0.93 \pm 0.25$	& $6.6 \pm 1.1$  	& --			\\
    QCD      & -- 			& $0.13 \pm 0.03$ 	& $0.16 \pm 0.04$  	& $18.0 \pm 2.9$	\\
    Total    & $19.1 \pm 1.1$ 		& $20.8 \pm 1.9$	& $41.3 \pm 2.4$  	& $23.3 \pm 3.0$	\\\hline
    Observed & $20$ 			& $10$			& $32$  		& $13$ \\
  \hline\hline
  \end{tabular}
  }
  \label{table:expectationsAbove300}

\end{table*}

\section{Systematic Uncertainties}\label{sec:SystematicSources}

The main source of systematic uncertainty results from the estimation of the background contributions that are dominated by the
statistical uncertainty of the data used in the control regions. These  uncertainties are in the range of 6 to 14\%. The
contamination from other backgrounds in these control regions has a negligible effect on the systematic uncertainty.  The
efficiencies for electron and muon reconstruction and identification are measured with the ``tag-and-probe"
method\,\cite{electronTrigger, muonTrigger} with a resulting uncertainty of 4.5\% for electrons and up to 3\% for muons. The
hadronic $\tau$ trigger efficiency is measured from $\Z\to\mutau$ events selected by single-muon triggers. This leads to a relative
uncertainty of 4.0\% and 6.4\% per $\tauhad$ candidate for the \tauelltauhad and \ditauhad final states,  respectively. Systematic
effects associated with \tauhad identification are extracted  from a fit to the \ZTT visible mass distribution,
$M(\tau_1,\tau_2)$. The fit constrains the \Z production cross section to the measured cross section  in \ZEE and \ZMM decay
channels, leading to a relative uncertainty of 6.8\% per $\tauhad$ candidate\,\cite{tauIDSyst}. The simulation is used to verify that
\tauhad identification efficiency remains constant as a function of \pt up to the high-mass signal region. 

Uncertainties that contribute to the $M(\tau_{1},\tau_{2},\MET)$  shape variations include the $\tauhad$ (2\%) and charged light-lepton
(1\%) energy scales, and the uncertainty on the \MET scale, which is used for the $M(\tau_{1},\tau_{2},\MET)$ mass calculation. The
\MET scale uncertainties contribute via the jet energy scale (2-5\% depending on $\eta$ and \pt) and unclustered energy scale
(10\%), where unclustered energy is defined as the energy not associated with the reconstructed leptons and jets with
$\pt >10\GeV$. The unclustered energy scale uncertainty has a negligible systematic effect on the signal acceptance and
$M(\tau_{1},\tau_{2},\MET)$ shape. In addition, the limited sizes of the simulated samples in the high-mass  regions lead to
systematic uncertainties in the background shape parametrization at high mass.  The uncertainty on the probability for a light quark
or gluon jet to be misidentified as a b jet (20\%) also has a negligible effect on the signal acceptance and
$M(\tau_{1},\tau_{2},\MET)$ mass shape.

The uncertainty on signal acceptance due to the PDF set included in the simulated samples is evaluated by comparing CTEQ6.6L,
MRST2006, and NNPDF10  PDF sets\,\cite{CTEQ,mrst2006,nnpdf10} with the default PDF set. The systematic effect due to imprecise
modeling of initial- and final-state radiation is determined  by reweighting events to account for effects such as missing $\alpha$
terms in the soft-collinear approach\,\cite{softCollinear} and missing  NLO terms in the parton shower
approach\,\cite{partonShower}. Finally, the uncertainty in the luminosity measurement is 2.2\%\,\cite{REFLUMI}.
Table~\ref{table:Systematics} summarizes the sources of  systematics  considered.

\begin{table*}[ht]
  \caption{Summary of the sources of systematic uncertainties.}
  \centering{
  \begin{tabular}{| l | c | c | c | c |}
  \hline\hline
       Source of Systematic             & \emu	& \etau & \mutau& \ditauhad     \\ [0.5ex] \hline
       Background estimation            & 7.4\% & 8.0\% & 5.8\% & 14.3\%      \\
       Muon id                          & 3.0\% & --    & 1.1\% & --	       \\
       Electron id                      & 4.5\% & 4.5\%	& --    & --	       \\
       Tau id	                        & -- 	& 6.8\% & 6.8\% & 9.6\%       \\
       Tau energy scale          	& -- 	& 2.1\% & 2.1\%	& 3.0\%       \\
       Tau trigger                      & -- 	& 4.0\%	& 4.0\%	& 9.0\%       \\
  \hline
       Jet energy scale      		& \multicolumn{4}{c|}{4.0\%}\\
       Luminosity                       & \multicolumn{4}{c|}{2.2\%}\\
       Parton distribution functions    & \multicolumn{4}{c|}{4.0 -- 6.5\%}\\
       Initial-state radiation          & \multicolumn{4}{c|}{3.1\%}	       \\
       Final-state radiation            & \multicolumn{4}{c|}{2.2\% }         \\
  \hline
  \hline
  \end{tabular}
  }
  \label{table:Systematics} 
\end{table*}

\section{Results}\label{sec:results}

The observed mass spectra shown in Fig.\,\ref{fig:masses} do not reveal any evidence for \zprimeditau production. A fit of the
expected mass spectrum to the data based on the \CLs criterion\,\cite{Read, Junk} is used to calculate an upper limit on the product
of the resonance cross section and its branching fraction into $\tau$-lepton pairs at 95\% CL as a function of the \zp mass for each
\ditau final state taking into account all systematic uncertainties shown in Table~\ref{table:Systematics}. The final limits are
obtained from the combination of all four final states.

\begin{figure}
  \begin{center}
    \includegraphics[angle=0,width=\cmsFigWidth]{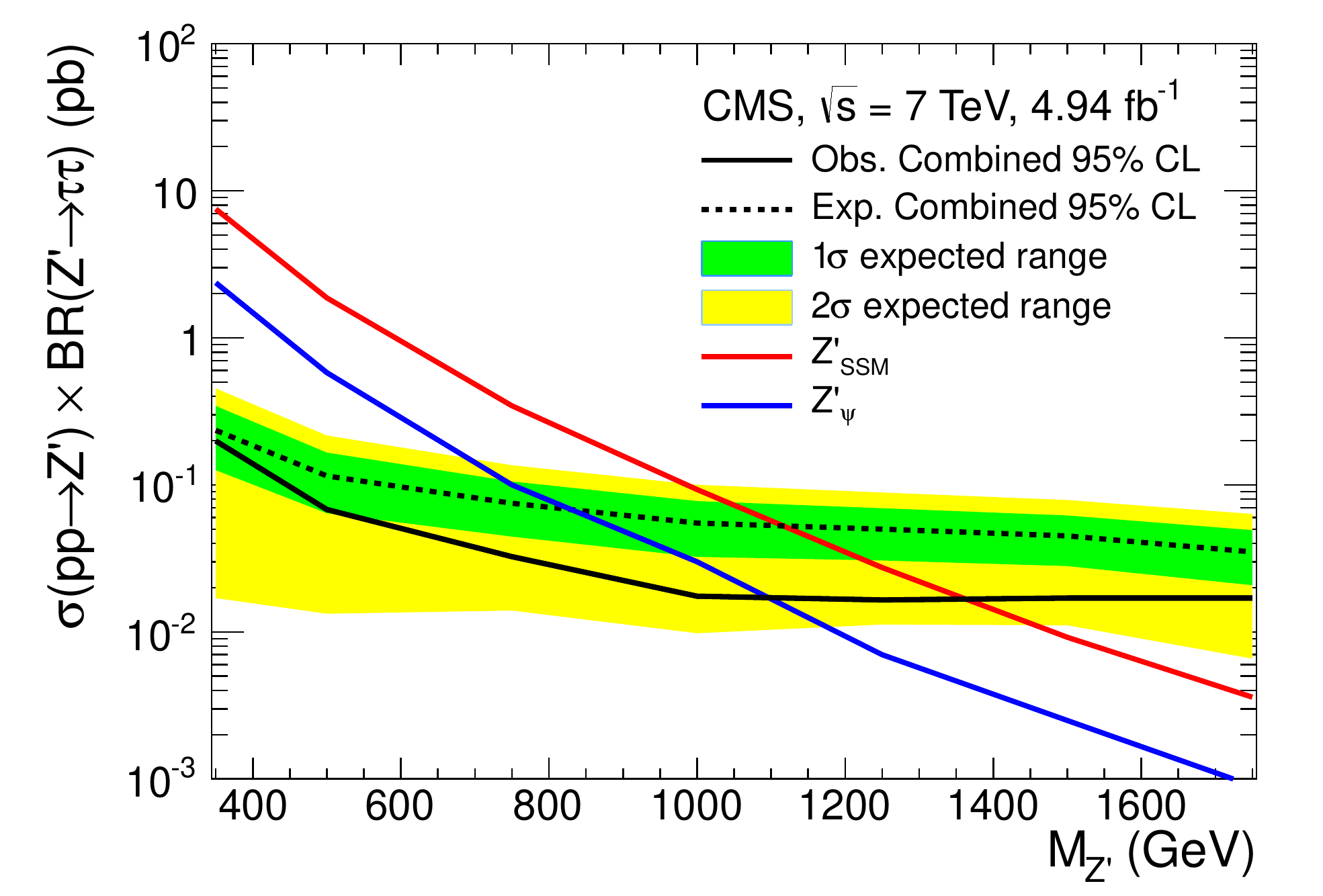}
    \caption{Combined upper limit at the 95\% CL on the product of the cross section and branching fraction into $\tau$ pairs as a
    function of the \zp mass. The bands represent the one and two standard deviations obtained from the background-only
    hypothesis.}
    \label{fig:Limits}
 \end{center}
\end{figure}

Figure\,\ref{fig:Limits} shows the combined expected and observed limits as well as the theoretical  cross section times the
branching fraction to $\tau$-lepton pairs for the SSM (\zprimeSSM) and E6 (\zprimePsi) models as functions of the \zp resonance
mass. The bands on the expected limits represent the one and two standard deviations obtained using a large sample  of
pseudo-experiments based on the background-only hypothesis. The upper limit on  $\sigma(\textrm{pp}\to\zp)\times
\textrm{Br}(\zp\to\ditau)$ corresponds to the point where the observed limit crosses the theoretical line.  We exclude a \zprimeSSM
with mass below 1.4\TeV and a \zprimePsi with mass less than 1.1\TeV. The \etau and \mutau final states contribute the most to the
limits. A downward fluctuation in the number of observed events with respect to the number of expected events leads to a limit that is
about 1.5 standard deviations higher than expected in the region above 600\GeV.

\section{Summary}\label{sec:conclusion}
A search for new heavy \zp bosons decaying into $\tau$-lepton pairs using data corresponding to an integrated luminosity of
$4.94 \pm 0.11\fbinv$ collected by the CMS detector in proton-proton collisions at $\sqrt{s}=7$\TeV was performed.  The observed
mass spectrum did not reveal any evidence for \zprimeditau production, and an upper limit, as a function of the \zp mass, on the
product of the resonance cross section and branching fraction into \ditau was calculated. The \zprimeSSM and \zprimePsi resonances
decaying to $\tau$-lepton pairs were excluded for masses below 1.4 and 1.1\TeV, respectively, at 95\% CL. These represent the
most stringent limits on the production of a new heavy resonance decaying into  $\tau$-lepton pairs published to date.

\section*{Acknowledgments}
We congratulate our colleagues in the CERN accelerator departments for the excellent performance of the LHC machine. We
thank the technical and administrative staff at CERN and other CMS institutes, and acknowledge support from: FMSR
(Austria); FNRS and FWO (Belgium); CNPq, CAPES, FAPERJ, and FAPESP (Brazil); MES (Bulgaria); CERN; CAS, MoST, and NSFC
(China); COLCIENCIAS (Colombia); MSES (Croatia); RPF (Cyprus); MoER, SF0690030s09 and ERDF (Estonia); Academy of
Finland, MEC, and HIP (Finland); CEA and CNRS/IN2P3 (France); BMBF, DFG, and HGF (Germany); GSRT (Greece); OTKA and NKTH
(Hungary); DAE and DST (India); IPM (Iran); SFI (Ireland); INFN (Italy); NRF and WCU (Korea); LAS (Lithuania);
CINVESTAV, CONACYT, SEP, and UASLP-FAI (Mexico); MSI (New Zealand); PAEC (Pakistan); MSHE and NSC (Poland); FCT
(Portugal); JINR (Armenia, Belarus, Georgia, Ukraine, Uzbekistan); MON, RosAtom, RAS and RFBR (Russia); MSTD (Serbia);
MICINN and CPAN (Spain); Swiss Funding Agencies (Switzerland); NSC (Taipei); TUBITAK and TAEK (Turkey); STFC (United
Kingdom); DOE and NSF (USA).

\bibliography{auto_generated}   

\cleardoublepage \appendix\section{The CMS Collaboration \label{app:collab}}\begin{sloppypar}\hyphenpenalty=5000\widowpenalty=500\clubpenalty=5000\input{EXO-11-031-authorlist.tex}\end{sloppypar}
\end{document}

%% file: EXO-11-031-authorlist.tex
\textbf{Yerevan Physics Institute,  Yerevan,  Armenia}\\*[0pt]
S.~Chatrchyan, V.~Khachatryan, A.M.~Sirunyan, A.~Tumasyan
\vskip\cmsinstskip
\textbf{Institut f\"{u}r Hochenergiephysik der OeAW,  Wien,  Austria}\\*[0pt]
W.~Adam, T.~Bergauer, M.~Dragicevic, J.~Er\"{o}, C.~Fabjan\cmsAuthorMark{1}, M.~Friedl, R.~Fr\"{u}hwirth\cmsAuthorMark{1}, V.M.~Ghete, J.~Hammer, N.~H\"{o}rmann, J.~Hrubec, M.~Jeitler\cmsAuthorMark{1}, W.~Kiesenhofer, V.~Kn\"{u}nz, M.~Krammer\cmsAuthorMark{1}, D.~Liko, I.~Mikulec, M.~Pernicka$^{\textrm{\dag}}$, B.~Rahbaran, C.~Rohringer, H.~Rohringer, R.~Sch\"{o}fbeck, J.~Strauss, A.~Taurok, P.~Wagner, W.~Waltenberger, G.~Walzel, E.~Widl, C.-E.~Wulz\cmsAuthorMark{1}
% ...
\vskip\cmsinstskip
\dag:~Deceased\\
1:~~Also at Vienna University of Technology, Vienna, Austria\\
% ...
